%Paper: chao-dyn/9403004
%From: Giovanni Gallavotti <giovanni@ipparco.roma1.infn.it>
%Date: Mon, 28 Mar 1994 11:00:19 +0100

%%%% This file is a Plain TeX file containing a few greek characters.
%%%% To use such characters you need to have the corresponding TFM
%%%% files (grreg10.tfm) and the PK (or equivalent) files at 300dpi.
%%%% Such public domain files, due to Silvio Levi, can be found in
%%%% various nodes.
%%%% If you HAVE such files set the parameter \greco=1; by default
%%%% the math fonts will be used as a (poor) surrogate.
%%%%%%%%%%%%%%%%%%%%%%%%%%%%%%%%%%%%%%%%%%%%%%%%%%%%%%%%%%%%%%%%%%%%%
\newcount\mgnf\newcount\tipi\newcount\tipoformule\newcount\greco

\greco=0         %=0 usa car. mat. per il greco, se no greco di Silvio Levi
\mgnf=0          %ingrandimento
\tipi=2          %uso caratteri: 2=cmcompleti, 1=cmparziali, 0=amparziali
\tipoformule=0   %=0 da numeroparagrafo.numeroformula; se no numero
                 %assoluto
%%%%%%%%%%%%%%%%%%%%%%%%%%%%%%%%%%%%%%%%%% INCIPIT
\ifnum\mgnf=0
   \magnification=\magstep0\hoffset=0.cm
   \voffset=-0.5truecm\hsize=16.5truecm\vsize=24.truecm
   \parindent=4.pt\fi
\ifnum\mgnf=1
   \magnification=\magstep1\hoffset=0.truecm
   \voffset=-0.5truecm\hsize=15.7truecm\vsize=24truecm
   \baselineskip=14truept plus0.1pt minus0.1pt \parindent=0.9truecm
   \lineskip=0.5truecm\lineskiplimit=0.1pt      \parskip=0.1pt plus1pt\fi
%%%%%%%%%%%%%%%%%%%%%%%%%%%%%%%%%%%%%%%%%%
%%%%%%%%%%%%%%%%%%%%%%%%%%% GRECO
%%%%%%%%%%%%%%%%%%%%%%%%%%%%%%%%%%%%%%%%%%%%%%%%
\let\a=\alpha \let\b=\beta       \let\d=\delta  \let\e=\varepsilon
         
\let\m=\mu                      \let\r=\rho
\let\s=\sigma \let\t=\tau

      \let\L=\Lambda  
     \let\F=\Phi

%%%%%%%%%%%%%%%%%%%%%%%%%%%%%%%%%%%%%%%%%%%%%%%%
%%%%%%%%%%%%%%%%%%%%%%%%%%%%%%%%%%%%%%%%%%%%%%%%%%%%%%%%%%%%%%
%%%%%%%%%%%%%%%%% EQUAZIONI CON NOMI SIMBOLICI
%%%
%%% per assegnare un nome simbolico ad una equazione basta
%%% scrivere \Eq(...) o, in \eqalignno, \eq(...) o,
%%% nelle appendici, \Eqa(...) o \eqa(...):
%%% dentro le parentesi e al posto dei ...
%%% si puo' scrivere qualsiasi commento;
%%% per assegnare un nome simbolico ad una figura, basta scrivere
%%% \geq(...); per avere i nomi
%%% simbolici segnati a sinistra delle formule e delle figure si deve
%%% dichiarare il documento come bozza, iniziando il testo con
%%% \BOZZA. Sinonimi \Eq,\EQ,\EQS; \eq,\eqs; \Eqa,\Eqas;\eqa,\eqas.
%%% All' inizio di ogni paragrafo si devono definire il
%%% numero del paragrafo e della prima formula dichiarando
%%% \numsec=... \numfor=...  (brevetto Eckmannn); all'inizio del lavoro
%%% bisogna porre \numfig=1 (il numero delle figure non contiene la sezione.
%%% Si possono citare formule o figure seguenti; le corrispondenze fra nomi
%%% simbolici e numeri effettivi sono memorizzate nel file \jobname.aux, che
%%% viene letto all'inizio, se gia' presente. E' possibile citare anche
%%% formule o figure che appaiono in altri file, purche' sia presente il
%%% corrispondente file .aux; basta includere all'inizio l'istruzione
%%%           \include{nomefile}
%%%%%%%%%%%%%%%%%%%%%%%%%%%%%%%%%%%%%%%%%%%%%%%%%%%%%%%%%%%%%%%

\global\newcount\numsec\global\newcount\numfor
\global\newcount\numapp\global\newcount\numcap
\global\newcount\numfig\global\newcount\numpag
\global\newcount\numnf

\def\SIA #1,#2,#3 {\senondefinito{#1#2}%
\expandafter\xdef\csname #1#2\endcsname{#3}\else
\write16{???? ma #1,#2 e' gia' stato definito !!!!} \fi}

\def \FU(#1)#2{\SIA fu,#1,#2 }

\def\etichetta(#1){(\veroparagrafo.\veraformula)%
\SIA e,#1,(\veroparagrafo.\veraformula) %
\global\advance\numfor by 1%
\write15{\string\FU (#1){\equ(#1)}}%
\write16{ EQ #1 ==> \equ(#1)  }}
\def\etichettaa(#1){(A\veraappendice.\veraformula)
 \SIA e,#1,(A\veraappendice.\veraformula)
 \global\advance\numfor by 1
 \write15{\string\FU (#1){\equ(#1)}}
 \write16{ EQ #1 ==> \equ(#1) }}
\def\getichetta(#1){Fig. \verafigura
 \SIA g,#1,{\verafigura}
 \global\advance\numfig by 1
 \write15{\string\FU (#1){\graf(#1)}}
 \write16{ Fig. #1 ==> \graf(#1) }}
\def\retichetta(#1){\numpag=\pgn\SIA r,#1,{\verapagina}
 \write15{\string\FU (#1){\rif(#1)}}
 \write16{\rif(#1) ha simbolo  #1  }}
\def\etichettan(#1){(n\verocapitolo.\veranformula)
 \SIA e,#1,(n\verocapitolo.\veranformula)
 \global\advance\numnf by 1
\write16{\equ(#1) <= #1  }}

\newdimen\gwidth
\gdef\profonditastruttura{\dp\strutbox}
\def\senondefinito#1{\expandafter\ifx\csname#1\endcsname\relax}
\def\BOZZA{
\def\alato(##1){
 {\vtop to \profonditastruttura{\baselineskip
 \profonditastruttura\vss
 \rlap{\kern-\hsize\kern-1.2truecm{$\scriptstyle##1$}}}}}
\def\galato(##1){ \gwidth=\hsize \divide\gwidth by 2
 {\vtop to \profonditastruttura{\baselineskip
 \profonditastruttura\vss
 \rlap{\kern-\gwidth\kern-1.2truecm{$\scriptstyle##1$}}}}}
\def\verapagina{
{\romannumeral\number\numcap}.\number\numsec.\number\numpag}}

\def\alato(#1){}
\def\galato(#1){}
\def\veroparagrafo{\number\numsec}\def\veraformula{\number\numfor}
\def\veraappendice{\number\numapp}
\def\verapagina{\number\pageno}\def\veranformula{\number\numnf}
\def\verafigura{{\romannumeral\number\numcap}.\number\numfig}
\def\verocapitolo{\number\numcap}\def\veranformula{\number\numnf}
\def\Eqn(#1){\eqno{\etichettan(#1)\alato(#1)}}
\def\eqn(#1){\etichettan(#1)\alato(#1)}

\def\Eq(#1){\eqno{\etichetta(#1)\alato(#1)}}
\def\eq(#1){\etichetta(#1)\alato(#1)}
\def\Eqa(#1){\eqno{\etichettaa(#1)\alato(#1)}}
\def\eqa(#1){\etichettaa(#1)\alato(#1)}
\def\dgraf(#1){\getichetta(#1)\galato(#1)}
\def\drif(#1){\retichetta(#1)}

\def\eqv(#1){\senondefinito{fu#1}$\clubsuit$#1\else\csname fu#1\endcsname\fi}
\def\equ(#1){\senondefinito{e#1}\eqv(#1)\else\csname e#1\endcsname\fi}
\def\graf(#1){\senondefinito{g#1}\eqv(#1)\else\csname g#1\endcsname\fi}
\def\rif(#1){\senondefinito{r#1}\eqv(#1)\else\csname r#1\endcsname\fi}
%%%%%%%%%%%%%%%%%%%%%%%%%%%%%%%%%%%%%%%%%%%%%%%%%%%%%%%%%%%%%%
%%%%%%%%%%%%%%%%%% Numerazione verso il futuro ed eventuali paragrafi
%%%%%%%%%%%%%%%%%% precedenti non inseriti nella scheda da compilare
%%%%%%%%%%%%%%%%%% e elenco referenze bibliografiche creato in
%%%%%%%%%%%%%%%%%% \jobname.bib
\def\bib[#1]{[#1]\numpag=\pgn
\write13{\string[#1],\verapagina}}

\def\include#1{
\openin13=#1.aux \ifeof13 \relax \else
\input #1.aux \closein13 \fi}

\openin14=\jobname.aux \ifeof14 \relax \else
\input \jobname.aux \closein14 \fi
\openout15=\jobname.aux%\write15
\openout13=\jobname.bib
%%%%%%%%%%%%%%%%%%%%%%%%%%%%

%%%%%%%%%%%%%%%%%%%%%%%%%%%%%%%%%%%%%%%%%%%%%%%%%%%%%%%%%%%%%%

\ifnum\tipoformule=1\let\Eq=\eqno\def\eq{}\let\Eqa=\eqno\def\eqa{}
\def\equ{}\fi

%%%%%%%%%%%%%%%%%%%%%%%%%%%%%%%%%%%%%%%%%%%%%%
%%%%%%%%%%%%%%%%%%%%%  Numerazione pagine

{\count255=\time\divide\count255 by 60 \xdef\hourmin{\number\count255}
	\multiply\count255 by-60\advance\count255 by\time
   \xdef\hourmin{\hourmin:\ifnum\count255<10 0\fi\the\count255}}

\def\oramin{\hourmin }

\def\data{\number\day/\ifcase\month\or gennaio \or febbraio \or marzo \or
aprile \or maggio \or giugno \or luglio \or agosto \or settembre
\or ottobre \or novembre \or dicembre \fi/\number\year;\ \oramin}

\setbox200\hbox{$\scriptscriptstyle \data $}

\newcount\pgn \pgn=1
\def\foglio{\number\numsec:\number\pgn
\global\advance\pgn by 1}
\def\foglioa{A\number\numsec:\number\pgn
\global\advance\pgn by 1}

\footline={\rlap{\hbox{\copy200}}\hss\tenrm\folio\hss}

%%%%%%%%%%%%%%%%%%%%%%%%%%%%%%%%%%%%%%%%%%%%% CARATTERI %%%%%%%%%%%%%%
\newskip\ttglue
%%cm semplificato
\def\TIPI{
\font\ottorm=cmr8   \font\ottoi=cmmi8
\font\ottosy=cmsy8  \font\ottobf=cmbx8
\font\ottott=cmtt8  %\font\ottosl=cmsl8
\font\ottoit=cmti8
%%%%% cambiamento di formato%%%%%%
\def \ottopunti{\def\rm{\fam0\ottorm}% passaggio a tipi da 8-punti
\textfont0=\ottorm  \textfont1=\ottoi
\textfont2=\ottosy  \textfont3=\ottoit
\textfont4=\ottott
\textfont\itfam=\ottoit  \def\it{\fam\itfam\ottoit}%
\textfont\ttfam=\ottott  \def\tt{\fam\ttfam\ottott}%
\textfont\bffam=\ottobf
\normalbaselineskip=9pt\normalbaselines\rm}
\let\nota=\ottopunti}
%%%%%%%%
%% am
\def\TIPIO{
\font\setterm=amr7 %\font\settei=ammi7
%\font\settesy=amsy7 \font\settebf=ambx7 %\font\setteit=amit7
%%%%% cambiamenti di formato %%%
\def \settepunti{\def\rm{\fam0\setterm}% passaggio a tipi da 7-punti
\textfont0=\setterm   %\textfont1=\settei
%\textfont2=\settesy   %\textfont3=\setteit
%\textfont\itfam=\setteit  \def\it{\fam\itfam\setteit}
%\textfont\bffam=\settebf  \def\bf{\fam\bffam\settebf}
\normalbaselineskip=9pt\normalbaselines\rm
}\let\nota=\settepunti}
%%%%%%%

%%cm completo
\def\TIPITOT{
\font\twelverm=cmr12
\font\twelvei=cmmi12
\font\twelvesy=cmsy10 scaled\magstep1
\font\twelveex=cmex10 scaled\magstep1
\font\twelveit=cmti12
\font\twelvett=cmtt12
\font\twelvebf=cmbx12
\font\twelvesl=cmsl12
\font\ninerm=cmr9
\font\ninesy=cmsy9
\font\eightrm=cmr8
\font\eighti=cmmi8
\font\eightsy=cmsy8
\font\eightbf=cmbx8
\font\eighttt=cmtt8
\font\eightsl=cmsl8
\font\eightit=cmti8
\font\sixrm=cmr6
\font\sixbf=cmbx6
\font\sixi=cmmi6
\font\sixsy=cmsy6
%%%%%%%%%%%%%%%%%%%%%%%%%%%%%%%%%%%%%%%
\font\twelvetruecmr=cmr10 scaled\magstep1
\font\twelvetruecmsy=cmsy10 scaled\magstep1
\font\tentruecmr=cmr10
\font\tentruecmsy=cmsy10
\font\eighttruecmr=cmr8
\font\eighttruecmsy=cmsy8
\font\seventruecmr=cmr7
\font\seventruecmsy=cmsy7
\font\sixtruecmr=cmr6
\font\sixtruecmsy=cmsy6
\font\fivetruecmr=cmr5
\font\fivetruecmsy=cmsy5
%%%% definizioni per 10pt %%%%%%%%
\textfont\truecmr=\tentruecmr
\scriptfont\truecmr=\seventruecmr
\scriptscriptfont\truecmr=\fivetruecmr
\textfont\truecmsy=\tentruecmsy
\scriptfont\truecmsy=\seventruecmsy
\scriptscriptfont\truecmr=\fivetruecmr
\scriptscriptfont\truecmsy=\fivetruecmsy
%%%%% cambio grandezza %%%%%%
\def \eightpoint{\def\rm{\fam0\eightrm}% switch to 8-point type
\textfont0=\eightrm \scriptfont0=\sixrm \scriptscriptfont0=\fiverm
\textfont1=\eighti \scriptfont1=\sixi   \scriptscriptfont1=\fivei
\textfont2=\eightsy \scriptfont2=\sixsy   \scriptscriptfont2=\fivesy
\textfont3=\tenex \scriptfont3=\tenex   \scriptscriptfont3=\tenex
\textfont\itfam=\eightit  \def\it{\fam\itfam\eightit}%
\textfont\slfam=\eightsl  \def\sl{\fam\slfam\eightsl}%
\textfont\ttfam=\eighttt  \def\tt{\fam\ttfam\eighttt}%
\textfont\bffam=\eightbf  \scriptfont\bffam=\sixbf
\scriptscriptfont\bffam=\fivebf  \def\bf{\fam\bffam\eightbf}%
\tt \ttglue=.5em plus.25em minus.15em
\setbox\strutbox=\hbox{\vrule height7pt depth2pt width0pt}%
\normalbaselineskip=9pt
\let\sc=\sixrm  \let\big=\eightbig  \normalbaselines\rm
\textfont\truecmr=\eighttruecmr
\scriptfont\truecmr=\sixtruecmr
\scriptscriptfont\truecmr=\fivetruecmr
\textfont\truecmsy=\eighttruecmsy
\scriptfont\truecmsy=\sixtruecmsy
}\let\nota=\eightpoint}

\newfam\msbfam   %per uso in \TIPITOT
\newfam\truecmr  %per uso in \TIPITOT
\newfam\truecmsy %per uso in \TIPITOT
%%%%%%%%%%%%%%%%%%%%%%%%%%%%%%%
%%Scelta dei caratteri
%\newcount\tipi \tipi=0   %e' definito all'inizio
\newskip\ttglue
\ifnum\tipi=0\TIPIO \else\ifnum\tipi=1 \TIPI\else \TIPITOT\fi\fi

%%%%%%%%%%%%%%%%%%%%%%%%%%%%%% DEFINIZIONI LOCALI
%%%%%%%%%%%%%%%%%%%%%%%%%%%%%%%%%%%%%%%%%%%%%%%%%%%%

\def\aps{{\it a posteriori}}
\let\0=\noindent\def\pagina{{\vfill\eject}}

\def\media#1{{\langle#1\rangle}}

\global\newcount\numpunt
\def\i#1{\def\9{#1}{\if\9.\global\numpunt=1\else\if\9,\global\numpunt=2\else
\if\9;\global\numpunt=3\else\if\9:\global\numpunt=4\ele
\if\9)\global\numpunt=5\else\if\9!\global\numpunt=6\else
\if\9?\global\numpunt=7\else\global\numpunt=8\fi\fi\fi\fi\fi\fi
\fi}\ifcase\numpunt\or{\accent18\char16.}\or{\accent18\char16,}\or
{\accent18\char16;}\or{\accent18\char16:}\or{\accent18\char16)}\or
{\accent18\char16!}\or{\accent18\char16?}\or{\accent18\char16\ \9}\else\fi}
%definisce la i con l' accento: i=\char16 e' la i senza punto
%e con lo spazio
%seguente corretto a secondo del carattere che segue
\def\XWPR{{\it a priori}}
\def\ap#1{\def\9{#1}{\if\9.\global\numpunt=1\else\if\9,\global\numpunt=2\else
\if\9;\global\numpunt=3\else\if\9:\global\numpunt=4\else
\if\9)\global\numpunt=5\else\if\9!\global\numpunt=6\else
\if\9?\global\numpunt=7\else\global\numpunt=8\fi\fi\fi\fi\fi\fi
\fi}\ifcase\numpunt\or{\XWPR.}\or{\XWPR,}\or
{\XWPR;}\or{\XWPR:}\or{\XWPR)}\or
{\XWPR!}\or{\XWPR?}\or{\XWPR\ \9}\else\fi}
% come \i ma per a priori
\def\XWPSR{{\it a posteriori}}
\def\aps#1{\def\9{#1}{\if\9.\global\numpunt=1\else\if\9,\global\numpunt=2\else
\if\9;\global\numpunt=3\else\if\9:\global\numpunt=4\else
\if\9)\global\numpunt=5\else\if\9!\global\numpunt=6\else
\if\9?\global\numpunt=7\else\global\numpunt=8\fi\fi\fi\fi\fi\fi
\fi}\ifcase\numpunt\or{\XWPSR.}\or{\XWPSR,}\or
{\XWPSR;}\or{\XWPSR:}\or{\XWPSR)}\or
{\XWPSR!}\or{\XWPSR?}\or{\XWPSR\ \9}\else\fi}
% come \i ma per a posteriori

\def\ie{\hbox{\it i.e.\ }}
\let\dpr=\partial\def\\{\hfill\break}

\def\*{\vglue0.3truecm}\let\0=\noindent
\let\io=\infty \def\V#1{\,\vec#1}
\let\ig=\int

\def\tende#1{\,\vtop{\ialign{##\crcr\rightarrowfill\crcr
              \noalign{\kern-1pt\nointerlineskip}
              \hskip3.pt${\scriptstyle #1}$\hskip3.pt\crcr}}\,}
\def\otto{\,{\kern-1.truept\leftarrow\kern-5.truept\to\kern-1.truept}\,}
\def\fra#1#2{{#1\over#2}}
\let\ciao=\bye

\let\==\equiv
\def\EE{{\cal E}}
%%%%%%%%%%%%%%%%%%%%%%%%%%%%%%%
%%%%%%%%%% GRECO di SILVIO LEVI

%\input d:/tipi/grmacro

%%%%%%%%
\def\ifnextchar#1#2#3{\let\tempe #1\def\tempa{#2}\def\tempb{#3}\futurelet
\tempc\ifnch}
\def\ifnch{\ifx\tempc\tempe\let\tempd\tempa\else\let\tempd\tempb\fi\tempd}
\def\gobble#1{}
\ifnum\greco=1\font\tengr=grreg10\else\font\tengr=cmmi10\fi
%\font\tengrbf=grbld10\font\tengrtt=grtt10
\def\greekmode{%
\catcode`\<=13
\catcode`\>=13
\catcode`\'=11
\catcode`\`=11
\catcode`\~=11
\catcode`\"=11
\catcode`\|=11
\lccode`\<=`\<%
\lccode`\>=`\>%
\lccode`\'=`\'%
\lccode`\`=`\`%
\lccode`\~=`\~%
\lccode`\"=`\"%
\lccode`\|=`\|%
\tengr\def\bf{\tengrbf}\def\tt{\tengrtt}}
\newcount\vwl
\newcount\acct
\def\lt{<}

{
  \greekmode
  \gdef>{\ifnextchar `{\expandafter\smoothgrave\gobble}{\char\lq\>}}
  \gdef<{\ifnextchar `{\expandafter\roughgrave\gobble}{\char\lq\<}}
  \gdef\smoothgrave#1{\acct=\rq137 \vwl=\lq#1 \dobreathinggrave}
  \gdef\roughgrave#1{\acct=\rq103 \vwl=\lq#1 \dobreathinggrave}
  \gdef\dobreathinggrave{\ifnum\vwl\lt\rq140	%if uppercase
    \char\the\acct\char\the\vwl\else\expandafter\testiota\fi}
      \gdef\testiota{\ifnextchar |{\addiota\doaccent\gobble}{\doaccent}}
        \gdef\addiota{\ifnum\vwl=\lq a\vwl=\rq370
            \else\ifnum\vwl=\lq h\vwl=\rq371 \else\vwl=\rq372 \fi\fi}
              \gdef\doaccent{\accent\the\acct \char\the\vwl\relax}
              }

\newif\ifgreek\greekfalse

\def\begingreek{\bgroup\greektrue\greekmode}
\def\endgreek{\egroup}

\let\math=$
{\catcode`\$=13
}

%%%%%%%%%%%%
\relax
\def\monos{\begingreek m'onos\endgreek}
\def\eidos{\begingreek e>~idos\endgreek}
\def\ortos{\begingreek >orj'os\endgreek}
\def\olos{\begingreek <'olos\endgreek}
\def\ergon{\begingreek >'ergon\endgreek}
\def\odos{\begingreek <od'os\endgreek}
\relax

\ifnum\greco=0

\def\odos{$\buildrel {\scriptscriptstyle\subset} \over o\delta \acute
o\varsigma$}

\def\olos{$\buildrel {\scriptscriptstyle\subset\prime} \over o\lambda
o\varsigma$}

\def\monos{$\mu \acute o\nu o \varsigma$}

\def\eidos{$\varepsilon\skew5\widetilde{\buildrel {\scriptscriptstyle\supset}
\over \iota}\delta o \varsigma$}

\def\ergon{$\buildrel {\scriptscriptstyle
\supset\prime} \over \varepsilon \rho\gamma o
\nu$}

\def\ortos{$\buildrel {\scriptscriptstyle \supset} \over o \rho\vartheta
\acute o\varsigma$}
\relax
\fi
\overfullrule=0pt

\def\eg{{\it e.g.\ }}
\vglue1.cm
\centerline{\bf Ergodicity, ensembles, irreversibility in Boltzmann
and beyond.
\footnote{${}^!$}{\nota expanded and revised version of a conference
read at the celebration of the $150^{th}$--anniversary of the birth of
Boltzmann, Vienna, 24 february, 1994.}}
\*
\centerline{\it Giovanni Gallavotti\footnote{${}^*$}{\nota
Dipartimento di Fisica, Universit\`a di Roma, P.le Moro 2, 00185
Roma. E--mail 40221::gallavotti. This paper is archived in
$mp\_arc@math.utexas.edu$, \#94-66 updated copies (in Postscript)
can also be obtained by sending request to the author, by E--mail.}}
\vglue1cm
{\bf Abstract:\it the implications of the original misunderstanding
of the etymology of the word "ergodic" are discussed, and the contents
of a not too well known paper by Boltzmann are critically examined.
The connection with the modern theory of Ruelle is attempted.}
\vglue1cm
{\bf\S1\it The etymology of the word "ergodic" and the heat theorems.}
\*\numsec=1\numfor=1
Trying to find the meaning of the word "ergodic" one is led to a 1884
paper by Boltzmann, [B84].\footnote{${}^1$}{\nota see the footnote of
S.  Brush in his edition, [Bo2], of the Lectures on Gas Theory, on p.
297 (\S32): here the Boltzmann's paper is quoted as the first place
where the word is introduced, although the etymology is taken from the
Erhenfests' paper, which is incorrect on this point: see [EE], note
\#93, p.89, (where also the first appearance of the word is incorrectly
dated and quoted).} This paper by Boltzmann is seldom quoted
\footnote{${}^2$}{\nota I found only the Brush's reference in ${}^1$,
and a partial account in [Br1], p.242 and p.  368, before my own
etymological discussion, appeared in print in [G1] after several years
of lectures on the subject.  My discussion was repeated in [G2] and
[G3].  More recently the paper has been appropriately quoted by [Pl],
unaware of my analysis.  The paper was discussed also by [Ma], see
footnote ${}^9$ below.} and no english translation is available yet.
But I think that this is one of the most interesting papers of
Boltzmann: it is a precursor of the work of Gibbs, [G], on the
ensembles, containing it almost entirely (if one recalls that the
equivalence of the canonical and microcanonical ensembles was already
established (elsewhere) by Boltzmann himself, at least in the free case
[B66],[B68]), and I will try to motivate such statement.

The paper stems from the fundamental, not too well known, work of
Helmholtz, [He1], [He2], who noted that {\it monocyclic}
systems\footnote{${}^3$}{\nota this is what we call today a system
whose phase space contains only periodic orbits, or cycles: \ie
essentially a one dimensional conservative system.} could be used to
provide models of thermodynamics in a sense that Boltzmann undertakes
to extend to a major generalization.

After an introduction, whose relative obscurity has been probably
responsible for the little attention this paper has received, Boltzmann
introduces the notion of "stationary" probability distribution on the
phase space of $N$ interacting particles enclosed in a vessel with
volume $V$.  He calls a family $\EE$ of such probabilities a {\it
monode}, generalizing an "analogous" concept on monocyclic
systems.\footnote{${}^4$}{\nota in fact Boltzmann first calls a monode
just a single stationary distribution regarded as an ensemble.  But
sometimes later he implicitly, or explicitly, thinks of a monode as a
collection of stationary distributions parameterized by some
parameters: the distinction is always very clear from the context.
Therefore, for simplicity, I take here the liberty of calling "monode"
a collection of stationary distributions, and the individual elements
of the collection will be called "elements of the monode".  The
etymology that follows, however, is more appropriate for the elements
of the monodes, as they are thought as consisting of many copies of the
same system in different configurations.  By reading the Boltzmann's
analysis one can get the impression, see p.  132 of [B84], that the
word monode had been already introduced by Maxwell, in [M]: however the
reference to Maxwell is probably meant to refer to the notion of
stationarity rather than to the word monode which does not seem to
appear in [M].}

In fact the orbits of a monocyclic system can be regarded as endowed
with a probability distribution giving an arc length a probability
proportional to the time spent on it by the motion: hence their family
forms a family of stationary probability distributions.

Etymologically this undoubtedly\footnote{${}^5$}{\nota of course one
can doubt (on this as well as on many other things).} means a family of
stationary distributions with a "unique nature", (each consisting of
systems with a "unique nature", differing only by the initial
conditions), from\ \monos\ and\ \ifnum\greco=0\eidos\else
\begingreek e>~idos\endgreek\fi,
with a probable reference to Plato and Leibnitz.

Then the following question is posed. Given an element $\m$ of a monode
$\EE$, also called a monode by Boltzmann, we can compute the average
values of various observables, \eg average kinetic energy, average total
energy, average momentum transfer per unit time and unit surface in the
collisions with the vessel walls, average volume occupied and density,
denoted, respectively:
$$T=\fra1N\media{K}_\m,\quad U=\media{K+\F}_\m,\quad p,\quad V,\quad
\r=\fra{N}V\Eq(1.1)$$
where $\F$ denotes the potential interaction energy and $K$ the total
kinetic energy. We then imagine to vary $\m$ in the monode $\EE$, by an
infinitesimal amount (this meanss changing any of the parameters
which determine the element). {\bf Question: \it is it true that the
corresponding variations $dU$ and $dV$ are such that}:
$$\fra{dU\,+\,p\,dV}T\quad{\rm is\ an\ exact\ differential}\quad dS\
?\Eq(1.2)$$
In other words is it true that the above quantities, defined in purely
mechanical terms, verify the same relation that would hold between them
if, for some thermodynamic system, they were the thermodynamic
quantities bearing the same name, with the further identification of
the average kinetic energy with the absolute
temperature?\footnote{${}^6$}{\nota that the temperature should be
identified with the average kinetic energy per particle was quite well
established (for free gases) since the paper by Clausius, [C], and the
paper on the equipartition of kinetic energy by Boltzmann, [B68] (in
the interacting cases); see the discussion of it in Maxwell's last
scientific work, [M].  The latter paper is also very interesting as
Maxwell asks there whether there are other stationary distributions on
the energy surface, and tries to answer the question by putting forward
the ergodic hypothesis.} If so the monode would provide a "mechanical
model of thermodynamics" extending, by far, the early examples of
Helmholtz on monocyclic systems.

Thus Boltzmann is led to the following definition:
\*
{\bf Definition: \it a monode $\EE$ is called an {\rm orthode} if the
property described by \equ(1.2) holds.}
\*
Undoubtedly the etymology of "orthode" is
\ \ortos\ and\ \ifnum\greco=0\eidos\else
\begingreek e>~idos\endgreek\fi, \ie "right nature".

I find it almost unbelievable that such a deep definition has not been
taken up by the subsequent literature. This is more so as Boltzmann, in
the same paper, proceeds to discuss "examples" of mechanical models of
thermodynamics, \ie examples of orthodic monodes.

It has, certainly, not escaped the reader that an orthodic monode (or orthode)
is what we call today an {\it equilibrium ensemble}. And the above
orthodicity concept is still attributed to Gibbs, see [Br1], p. 242).

The examples of orthodes discussed by Boltzmann in his paper are
the {\it holode} and the {\it ergode} which are two ensembles whose
elements are parameterized with two parameters $\b,N$ or $U,N$,
respectively. Their elements are:
$$\m_{\b,N}(d\V pd\V q)=\fra{d\V p_1\ldots d\V p_nd\V q_1\ldots d\V
q_n}{const} \,e^{-\b(K+\F)}\Eq(1.3)$$
and:
$$\m_{U,N}(d\V p d\V q)=\fra{d\V p_1\ldots d\V p_nd\V q_1\ldots d\V
q_n}{const} \,\d(K(\V p)+\F(\V q)-U)\Eq(1.4)$$
{\it Boltzmann proves that the above two ensembles are both orthodes!}
thus establishing that the canonical and the microcanonical ensembles
(using our modern terminology) are equilibrium ensembles and provide
mechanical models of thermodynamics.\footnote{${}^7$}{\nota he also
studies other ensembles, for instance in a system in which angular
momntum is conserved, \eg a gas in a spherical container, he considers
the stationary distributions with fixed energy and fixed total angular
momentum $\V L$.  Such monodes are called, by Boltzmann, {\it planodes}
(form the "area law"); and he remarks that in general they are not
orthodic (in fact one needs the extra condition that $\V L=\V0$).}

Boltzmann's proof makes use of the auxiliary (with respect to the above
definition) notion of heat transfer: in the canonical case it yields
exactly the desired result; in the microcanonical it is also very
simple but somehow based on a different notion of heat transfer.  An
analysis of the matter easily shows, [G4], that the
correct\footnote{${}^8$}{\nota there is a problem only if one insists
in defining in the same way the notion of heat transfer in the two
cases: this is a problem that Boltzmann does not even mention, possibly
because he saw as obvious that the two notion would become equivalent
in the thermodynamic limit.} statement becomes exact only in the limit
as $N,U\to \io$, keeping of course $\fra{U}V,\fra{N}V$ constant, \ie in
what we call today the "thermodynamic limit".

Undoubtedly the word "holode" has the etymological origin of
\ \olos\ and\ \ifnum\greco=0\eidos\else
\begingreek e>~idos\endgreek\fi\ while
"ergode" is a shorthand for "ergomonode" and it has the etymological
root of\ \ergon\ and\ \ifnum\greco=0\eidos\else
\begingreek e>~idos\endgreek\fi, meaning a "monode with given energy", [G1].
\footnote{${}^9$}{\nota the word "ergode" appears for the first time on
p.  132 of [B84]: but this must be a curious misprint as the concept is
really introduced on p.  134.  On p.  132 the Author probably meant to
say "holode", instead: this has been correctly remarked by [Pl].  See
also footnote 13.  The above etymology was probably proposed for the
first time by myself in various lectures in Roma, and it was included
in the first section of [G1].  The date of the preprint of [G1] is june
1980, the publication date is 1981: a year later a reference to the
same new etymology appears, see [Ja],[Ma], attributed to Mathieu.  I
find it obviously possible, even likely, that independently two
scientists may reach the same conclusion: even with only a few years of
delay.  Nevertheless {\it no reference} is
made to my book in the paper of 1988 by Mathieu, in [Ma].  In fact I
gave a series of lectures in august 1979 in Cortona which were attended
by prof.  R.  Nagel who had access to (and, as all the partecipants, a
copy of) my manuscript [G1] already including the etymology section in
its present form; he informed me in a subsequent letter that he had
discussed the matter with his student Mathieu, sending me a manuscript
by him on the subject.}.  The word "holode" is probably a shorthand for
"holomonode", meaning a "global monode" (perhaps a monode involving
states with arbitrary energy, \ie spread over the whole phase space).

This is not what is usually believed to be the etymology of "ergode":
the usual belief comes from the Erhenfests' statement that the
etymology is\ \ergon\ and\ \odos,
with the meaning of "unique path on the surface of
constant energy, see [EE] note \#93.  This absurd etymology has been
taken up universally and has been attached to the subject of "ergodic
theory", which is instead a theory dealing with time evolution properties.

\*
{\bf\S2 \it The ergodic hypothesis, continuous and discrete phase
space.}
\*\numsec=2\numfor=1
The etymological error of the Erhenfests could be just an amusing
fact: but it had a rather deep negative influence in the development of
the 20-th century Physics.  They present their etymology in connection
with the discussion (amounting to a {\it de facto} rejection) of the ergodic
hypothesis of Boltzmann.  In fact Boltzmann had come to the ergodic
hyptothesis in his attempts to justify, \ap, that the ergode, as a model of
thermodynamics, had to produce $\underline{\bf the}$
thermodynamics of a system
with the given hamiltonian function, (and not just a model).

Boltzmann had argued that the trajectory of any initial datum
evolves on the surface of constant energy, visiting all phase space
points and spending equal fractions of time in regions of equal
Liouville measure.

The Ehrenfests criticize such a viewpoint on surprisingly abstract
mathematical grounds: basically they say that one can attach to each
different trajectory a different label, say a real number, thus
constructing a function on phase space constant on trajectories. Such a
function would of course have to have the same value on points on the
same trajectory (\ie it would be a constant of motion). This is stated in
the note \#74, p. 86  where the number of different paths is even
"counted", and referred to in the note \#94, p. 89. Therefore, they
conclude, it is impossible that there is a single path on the surface of
constant energy, \ie the ergodic hypothesis is inconsistent (except for
the monocyclic systems, for which it trivially
holds).\footnote{${}^{10}$}{\nota the abstract
mathematical nature of this argument, see also below for a critique, was
apparently remarked only by a mathematician, see [Pl] p. 86, althoug a
great one (Borel, 1914); but it escaped many physicists. It is worrying
to note how seriously the mathematicians took the ergodic hypothesis and
how easily they disposed of it, taking for granted that the Ehrenfests
formulation was the original formulation by Boltzmann and Maxwell, see
[Br1], p. 383.}

Having disposed of the ergodic hypothesis of Boltzmann, the Erhenfests
proceed to formulate a new hypothesis, the rather obscure (and somewhat
vague as no mention is made to the frequency of visit to regions in
phase space) "quasi ergodic hypothesis", see notes \#98 and \#99, p.90, in
[EE]: it led the physicists away from the subject and it inspired the
mathematicians to find the appropriate definition giving birth to
ergodic theory and to its first non trivial results.

The modern notion of ergodicity is not the quasi ergodicity of the
Erhenfests. It is simply based on the remark that the Erhenfests had
defined a non trivial constant of motion very abstractly, by using the
axiom of choice.  In fact from the definition, consisting in attaching
a different
number, or even $6N-2$ different numbers, to each distinct trajectory,
there is {\it in principle} no way to construct a table of the values
of the function defined in order to distinguish the different
trajectories.  In a system ergodic in the modern sense the Ehrenfests'
construction would lead to a non measurable function; and to a
physicist dowed with common sense {\it such a function, which in
principle cannot be tabulated, should appear as non existent, or as non
interesting}.  Thus the motion on the energy surface is called ergodic
if there are no {\it measurable} constants of motion: here measurable
is a mathematical notion which essentially states the possibility of a
tabulation of the function.

It is surprising that a generation of physicists could be influenced
(in believing that the ergodic hypothesis of Boltzmann had to be
abandoned as a too naive viewpoint) by an argument of such an
exquisitely abstract nature, resting on the properties of a function
that could not be tabulated (and not even defined if one did not accept
the sinister axiom of choice).\footnote{${}^{11}$}{\nota we recall,
as it is quite an irony, the coincidence that the
recognition and the development of the axiom of choice was due
essentially to the same Zermelo who was one of the strongest opponents
of Boltzmann ideas on irreversibility, see also [Sc].}

Therefore it is worth, perhaps, to try understanding what could have
possibly meant Boltzmann when he formulated the ergodic hypothesis.
Here one cannot fully rely on published work, as the question was never
really directly addressed by Boltzmann in a critical fashion (he might
have thought, rightly, that what he was saying was clear enough).
The following analysis is an elaboration of [G1], [G2]: in some points
it gets quite close to [Pl].  It will not escape the reader that [Pl] has
a somewhat different point of view on several key issues, although we
seem to share the main thesis that the [EE] paper is responsible for most
of the still persisting misunderstandings on Boltzmann's work. Including
the exclusive attribution to Gibbs of Boltzmann's ideas on ensembles, so
clearly elaborated in [B84].

My point of view is that of those who believe that Boltzmann always
conceived the phase space as a discrete space, divided into small
cells, see [B72],  p. 346.  He always stressed that the continuum must be
understood as a limit, see [Br], p. 371, and [Kl1,2,3],[D]. The book of
Dugas, [D], is particularly illuminating (also) on this respect (see
for instance ch. 1 and the quotations of Boltzmann presented there,
where he seems to identify the discrete viewpoint with the atomistic
conceptions).

Although Boltzmann seems to have been, sometimes, quite apologetic about
such a viewpoint (even calling it a "mathematical fiction", [Ba], p.18,
from [B72]; see also [Pl], p. 75), he took advantage of it to a
point that one can say that most of his arguments are based on a
discrete conception of phase space, followed at the end by a passage to
the continuum limit.  It should be however understood that the
discretization that Boltzmann had in mind is by no means to be
identified with the later concept of coarse graining: see \S4 where a
modern version of Boltzmann's discretization is considered and where a
distinction has to be made between cells and volume elements, see also
[Pl] and [G3].

It is easier for us, by now used to numerical simulations, to
grasp the meaning of a cell: in the numerical simulations a cell is
nothing else but an element of the discrete set of points in phase
space, each represented within computer precision (which is finite). One
should always discuss how much the apparently harmless discreteness of
the phase space affects the results. This is, however, almost never
attempted: see [G3] for an attempt. A volume element has, instead, a
size much larger than the machine resolution, so that it looks a
continuum (for some purposes).

Hence one can say that an essential characteristics of Boltzmann's
thought is to have regarded a system of $N$ atoms, or molecules, as
described by a {\it cell} of dimension $\d x$ and $\d p$ in each
position and momentum coordinates. He always proceeded by regarding such
quantities as very small, avoiding to enter into the analysis of their
size, but every time this had some importance he must have regarded them
as positive quantities.

A proof of this is when he refutes the Zermelo's paradoxes by counting
the number of cells of the energy surface of $1cm^3$ of normal air,
[B96], a feat that can only be achieved if one considers the phase
space as discrete.

In particular this point of view must have been taken when he
formulated the ergodic hypothesis: in fact conceiving the energy
surface as discrete makes it possible to assume that the motion on it
is "ergodic", \ie it visits {\it all} the phase space points, compatible
with the given energy (and possibly with other "trivial" constants of motion)
behaving as a monocyclic system (as all the motions are necessarily
periodic).

The passage to the continuum limit, which seems to have never been made by
Boltzmann, of such an assumption is of course extremely delicate, and
it does not lead necessarily to the interpretation given by the
Erhenfests.  It can easily lead to other interpretations, among which
the modern notion of ergodicity: but it should not be attempted here, as
Boltzmann himself did not attempt it.

And in general one can hardly conceive that by studying the continuum
problem could lead to really new information, that cannot be obtained
by taking a discrete viewpoint.  Of course some problems might still be
easier if studied in the continuum, [S]: and the few results on
ergodicity of physical systems do in fact rely explicitly on continuum
models.  However I interpret such results rather as illustrations of
the complex nature of the discrete model: for instance the ergodicity
theory of a system like a billiards is very enlightening as it allows us
to get some ideas on the question of whether there exist other ergodic
distributions (in the sense of ergodic theory) on the energy surface,
and which is their meaning, [BSC].

And the theory of the continuum models has been essential in providing
new insights in the description of non equilibrium phenomena, [R],
[CELS].

Finally the fruitfulness of the discrete models can be even more
appreciated if one notes that they have been the origin of the quantum
theory of radiation: it can be even maintained that already Boltzmann had
obtained the Bose Einstein statistics, [Ba].

The latter is a somewhat strong intepretation of the 1877 paper, [B77].
The most attentive readers of Boltzmann have, in fact, noted that in his
discretizations he really thinks always in terms of the continuum limit
as he does not discuss the two main "errors" that one commits in
regarding a continuum formulation as an approximation (based on
integrals instead of sums)\footnote{${}^{12}$}{\nota and which amount to
the identification of the Maxwell Boltzmann statistics and the Bose
Einstein statistics, and to neglecting the variation of physically
relevant quantities over the cells: see the lucid analysis in [K], p.60;
for a technical discussion see [G3],[G4].} with respect to a discrete
one.

The above "oversight" might simply be a proof that Boltzmann never took
the discretization viewpoint to its extreme consequences.  Among which
there is that the equilibrium ensembles are {\it no longer} orthodic in
the sense of Boltzmann (see [G3],[G4]), (although they still provide a
model for thermodynamics provided the temperature is no longer
identified with the average kinetic energy): a remark that very likely
was not made by Boltzmann {\it in spite of his consideration and
interest on the possibility of finding other integrating factors for
the heat transfer $d Q$}, see the footnote on p.  152 in
[B84].\footnote{${}^{13}$}{\nota I have profited, in checking my
understanding of the original paper as partially exposed in [G1], from
an english translation that Dr.  J.  Renn kindly provided, while being
ny student in Roma (1984).  I could note this footnote in [B94], and
insert a few new remarks in the present paper, because of his
translation, (unfortunately still unpublished).}

The necessity of an understanding of this "oversight" has been in
particular clearly advocated by Kuhn referring to Boltzmann's "little
studied views about the relation between the continuum and the
discrete", [K], for instance.
\*
{\bf\S3 \it The ergodic hypothesis and irreversibility.}
\*\numsec=3\numfor=1
The reaction of the scientific world to the ergodic hypothesis was, "on
the average", a violently negative one, also as it was intended to
provide further justification to the irreversibility predicted by the
Boltzmann equation, derived earlier.

The great majority of the scientists  saw absurd and paradoxical
consequences of the hypothesis, without apparently giving any importance
to the "unbelievable" fact that on the basis of a maximal simplicity
assumption (\ie only one cycle on the energy surface) Boltzmann was
obtaining not only the possibility of explaining, mechanically, the
classical equilibrium thermodynamics but also that of explaining it in a
quantitative way. It allowed, for the first time, the theoretical
calculation of the equations of state of many substances (at least in
principle) like imperfect gases, and even other fluids and solids.

The success of the highly symbolic but very suggestive
formula of Boltzmann, see [EE], p.25:
$$\lim_{T\to\io}\fra{dt}T=\fra{\s\,ds}{\ig\s\,ds}\Eq(3.1)$$
(where $\s$ is the microcanonical density on the energy surface, whose
area element is $ds$) in the calculation of the equilibrium properties
of matter led quickly the physicists to accept it in the "minimal
interpretation".  Such interpretation demanded that the r.h.s.  be used
to compute the equilibrium averages and the l.h.s.  ignored, togheter
with the atomic hypothesis.  This is regarded as a {\it law of nature},
in spite of the persistent skepticism (or deep doubts) on its
deducibility from the laws of mechanics.  A point of view
usually attributed to Gibbs, referring to [G], and which is still around
us, although we assit, since the mid fifties, to a slow but inexorable
inversion of tendency.

Immediately after the first critiques Boltzmann elaborated answers
often very clear and simple by our modern understanding: but they were
very frequently ill understood not only by the opponents of Boltzmann
and their epigones, but also by those who were closest to him. The
above quoted critique to the ergodic hypothesis by the Erhenfests is a
shocking example.

Another example is the recurrence paradox, based on the simple theorem
of Poincar\'e. Boltzmann was finally led to the calculation of the
number of cells on the energy surface, [B96], thus to a
superastronomical estimate of the recurrence time: which, nevertheless,
did not seem to impress many.

It is also clear that Boltzmann himself became aware of the
fact that, after all, the ergodic hypothesis might have been
unnecessarily strong and perhaps even useless to explain the approach to
equilibrium in physical systems. The latter in fact reach equilibrium,
normally, within times which are microscopic times, not at all
comparable with the recurrence time. He asserted repeatedly that the
(very few) macroscopic observables of interest had essentially the {\it
same} value in most of the energy surface, and the time spent in the
"anomalous phase space cells" is therefore extremely small: a
quantitative understanding of this is provided by the Boltzmann
equation. This remark also frees \equ(3.1) from the ergodic hypothesis: it
might well be that the r.h.s can be used to evaluate the average values,
in equilibrium, of the few observables which are of interest, although
there might be observables (\ie functions on phase space) for which the
\equ(3.1) fails.

It is well known that Boltzmann went quite far in this direction, by
providing us with a concrete method to estimate the  true
times of approach to
equilibrium: the Boltzmann's equation (historically developed well
before the 80's).

Finally it is worth noting that the methods used by By Boltzmann in
deriving the theory of the ensembles and the ergodic hypothesis
are quite modern and in fact are
most suited to illustrate the new developments on non equilibrium
theory: as I shall try to prove in the next section.
\*
{\bf\S4 \it Non equilibrium. Ruelle's principle. Outlook.}
\*\numsec=4\numfor=1
I cannot resist the temptation of at least mentioning some recent new
developments which look to me exciting and very likely to remain as
important progress in the field.\footnote{${}^{14}$}{\nota I like to think
that Boltzmann his listening to the celebration of his birthday: he
would certainly be bored by hearing a, presumably poor, exposition
dealing only with things that he knew far better.}

The \equ(3.1), in its minimal interpretation of providing, via the r.h.s.
(\ie the microcanonical distribution), the law for the evaluation of the
"relevant" macroscopic observables, starting from the energy function of
the system, "solves" the problem of the equilibrium theory. Completely,
as far as we know (in Classical Physics).

Is a similar theory possible for systems in non equilibrium, but in
a stationary state? What (if anything) replaces the microcanonical
distribution in such cases? As an example of "cases" we mean the motion
of a gas of particles subject to a constant force ("electric field")
setting them in motion, while the energy produced is dissipated into a
reservoir.

The answer seems positive, at least in some cases.  The problem lies
in the fact that the motion of such systems is dissipative, hence the
volume element of the energy surface is not conserved even in the
simple case in which the thermostat is such that it keeps the total
energy of the system constant (as I shall suppose, to simplify the
discussion), \ie the microcanonical distribution cannot describe the
stationary state.  Taking the continuum viewpoint we can imagine that
the motion is essentially concentrated, after a transient time, on a
set $A$ which has zero measure with respect to the Liouville measure on
the energy surface.

To avoid giving the impression that the discussion is abstract (hence
possibly empty) let me declare explicitly one, among many, models that
one should have in mind. We consider a system of $N$ particles
interacting with a potential energy $\F$ and subject to an external
constant force field $\V E$, (\eg electric field):
$$\dot{\V q_i}=\fra1m{\V p_i},
\qquad\dot{\V p_i}=-\dpr_{\V q_i}\F+\V E-\a(\V p)\V p_i\Eq(4.1)$$
where $\V E$ is the external constant force and $\a$ is defined so that
the energy $\sum_{i=1}^N\fra{\V p_i^2}{2m}+\F$ is constant (\ie
$\a=\fra{\V E\cdot\sum\V p_i}{\sum \V p_i^2}$). The term $\a\V p_i$ is a
model of a thermostat (this should be called a {\it gaussian
thermostat} as it is related to the Gauss' principle of "least
constraint", see [CELS]). The system is considered enclosed in a box with
periodic boundary conditions: hence we expect that a current parallel to
$\V E$ will be established and the system will reach a stationary state.
The volume in phase space contracts at a rate $(3N-1)\a$, (which is
positive, in the average): hence the motion will asymptotically develop
on some "attractor", which is a set of $0$ Liouville measure.

What follows will lead to a unified theory of the equilibrium as well as the
non equilibrium, for system \equ(4.1).

The discrete viewpoint is also possible: the enegy surface consists of
cells which are relevant (for the study of the asymptotic properties)
forming a set $A$ in phase space, and of cells which are irrelevant.
The motion can be regarded to develop on the set of cells which are in
$A$, which is strictly smaller than the set of all the cells: in fact
far smaller (and in the continuum limit the fraction of cells in $A$
approaches $0$).

Since the volume of the cells is not conserved care must be exercised
in regarding the dynamics as a permutation of the cells of $A$.  This
is in fact also true in the equilibrium case because, even if the cells
do not change in volume, they are deformed being squeezed in some
directions and dilated in others.  In equilibrium it is possible to
conceive situations in which the deformation can be neglected (this
leads to restrictions on the region of temperature and density in which
the consideration of the dynamics as a cell permutation is acceptable:
a discussion which we have not begun above and which we avoid here as
well, see [G3] for a quantitative analysis). And a similar analysis can
be carried in the present case.

Basically one has to think that the system is observed at time
intervals $\t_0$ which are not too small (so that something really
happens) and not too large (so that the cell's deformations can be
either neglected or controlled, at least for a large majority of
cells): see [G3] for a quantitative analysis of what this means in the
equilibrium cases and of when this might lead to inconsistencies.  Let
$S_{\t_0}$ denote the transformation of $A$ describing the dynamics on
$A$ over the time $\t_0$.  By making the cells small enough we can take
$\t_0$ larger.

We shall imagine the set $A$ as a surface in phase space of dimension
roughly $\fra{6N}2$ at least if the external force is small (so that
the friction $\a$, \ie the phase space volume contraction, is also
small): in fact if there is no external force the dimension of $A$
should be $1+\fra{6N-2}2$.\footnote{${}^{15}$}{\nota because there are
as many contracting directions as expanding ones (the volume being
conserved in the $6N$ dimensional phase space); and there are two
"neutral" directions (the direction orthogonal to the energy surface
and the direction of the phase space motion) one of which lies on the
energy surface (the direction of motion), see [Dr], [ECM1], [SEM].  Of course
the existence of other conserved quantities, as in \equ(4.1) when the
linear momentum is conserved, affects this calculation: in \equ(4.1),
when $\V E=\V0$,
this brings down the dimension to $1+\fra{6N-8}2$.  Furthermore we are
assuming here that there are no "neutral" directions other than the
ones possibly provided by the obvious conservation laws: \ie that our
system has strong instability properties (hence this does not {\it
directly} apply to the free gas, for instance).} The surface $A$ can fold
itself on the energy surface filling it up completely (in the $\V
E=\V0$ case) or not (in the general case).\footnote{${}^{16}$}{\nota in
the continuum point of view we can proceed as follows: we fix an
approximation $\e$ and we identify the points on $A$ which are very far
on any path that joins them {\it along} $A$, but which are close within
$\e$ as points on the energy surface.  Then $A$ becomes a finite
surface $A_\e$.  This surface depends on the point that we initially
choose for the construction: but the results should be independent on
the choice.  The latter is in fact an assumption which essentially
replaces the ergodicity assumption of the conservative cases.  The
above "viewpoint" will imply ergodicity in the case of the conservative
systems: this non trivial fact is a consequence of the hidden
assumption that the description does not depend on which surface $A_\e$
we choose as an approximation for $A$.  In fact the choice of $A_\e$
suffers from an arbitrariness which consists in deciding that one given
point is actually on $A_\e$: choosing another point leads, in general,
to a different $A_\e$.  In concrete cases it will, however, be very
difficult to show that the results are independent on $A_\e$ (a
manifestation of the conservation of difficulties).} We can assume the
following extension of the ergodic hypothesis: {\it on $A$ the dynamics
is a one cycle permutation of the cells}.

Then the motion of a randomly chosen initial datum, randomly with
respect to a distribution with some density on the energy surface, will
simply consist in a fast approach to the surface $A$; at the same time
data which are on $A$ itself and close to each other will separate from
each other at some exponential rate, because on $A$ all the directions
are dilated, by definition.  To fix the ideas we take the initial data
with constant density in some little ball $U$.  If we assume, for
simplicity, the above ergodic hypothesis, the layer is, over times
multiples of the recurrence time, a set of cells each visited with
equal frequency.  However the surface $A$ will, in general, not be a
monolayer of cells but it will have a large "width", \ie a
(macrospcopic) area element $d\s$ will contain many (microscopic)
cells.  \footnote{${}^{17}$}{\nota this can perhaps be clarified if one
thinks of the numerical experiments in which the computer
representatives of the phase space points are regarded as cells, while
the unstable manifolds of the motion are regarded as surfaces built
with computer points, \ie cells.}

The number of cells per unit area can be deduced by remarking that after
a time $\t=M\t_0$ the density of cells around $x\in A$, initially
distributed with constant density in the region $U$ (where the initial
data are randomly chosen), has to be proportional to the inverse of the
area expansion rate of the transformation $S_\t$. This means that
we expect that the distribution on $A$ which has to be used to compute
the stationary averages is described by a suitable density with respect
to the area element on $A$.

With this intuitive picture in mind, [R], ECM2], we see that a little
ball $U$ in phase space evolves becoming a thin layer around $A$: the
density of the layer, after a large time $T$, is proportional to the
expansion rate of the surface area on $A$ under the transformation
$S_T$ generating the time evolution over the given time.

In the case of no external forces one has that the surface $A$ folds
itself on the energy surface coming back to a given phase space volume
element $V_0$ (not to be confused with a cell, which has to be thought
as much smaller); just enough times, and with enough volume around, so
that the fraction of the volume initially in $U$ and falling in the
volume element $V_0$ is proportional to $V_0$ itself (this is
consistent because of the equality of the total expansion rate and the
total contraction rate, due to the hamiltonian nature of the equations
of motion).  But in general the fraction of volume $U$ falling into a
volume element will be far different from the volume element fraction
of the energy surface.

One is thus led to the following unified "principle" to describe the
stationary states of non equilibrium systems, [R]:
\*
{\bf Principle: \it the average values of the observables in the
stationary state describing the asymptotic behaviour of systems like
\equ(4.1), is computable from a probability distribution on $A$ which
has a density, with respect to the surface element of
$A$.}\footnote{${}^{18}$}{\nota it is extremely important to think, to
avoid trivial contradictions, that the cells on $A$ must be regarded as
much smaller than the surface elements of $A$ that we consider in
talking about the density.}
\*
This principle can be more mathematically stated (a problem into which
we refrain to enter here), and is due to Ruelle, [R], who based himself
also on the results of Anosov, Sinai, Bowen on the theory of a class of
dynamical systems known as "hyperbolic systems" (which play in some
sense, for non equilibrium statistical mechanics, the role of the
monocyclic systems of Helmholtz). The probability distributions selected
by the above principle (which in "good cases" is unique) are called SRB
measure, [R].

What is the predictive value of the above statements? in the cases
without external forces we have already mentioned that this principle
leads to the microcanonical distribution and, therefore, implies the
classical thermodynamics, [B84].  Life is made easy by the fact that
although $A$ may be very difficult to identify, still the stationary
distribution is just the microcanonical ensemble because $A$ folds on
the energy surface filling it up completely, with no gaps.

In the dissipative cases it seems that we have little control on $A$
and hence on the stationary distribution.

{\it Yet this might not be really so:} we simply have to learn how to
extract informations from such an abstract principle.  After all it now
seems natural that the Gibbs distribution predicts all the phenomena of
equilibrium statistical mechanics (from the phase coexistence, to the
critical point, to cristallization).  But this was far from clear only
a few decades ago, and many decades after the original formulations of
Maxwell, Gibbs and Boltzmann (as many of us certainly recall).

That the principle might have predictive value is indicated by the
first attempts at its use in problems of statistical mechanics, see
[ECM2], (see also [CELS]), who were somewhat inspired by previous
papers, see also [HHP].  In fact only recently the principle
started being considered in the theory of non equilibrium, as it was
developed originally by Ruelle mainly as an attempt to a theory of
turbulent phenomena. This is not the appropriate place to discuss the
paper [ECM2] in the perspective of the above principle: the discussion
is rather delicate (as [ECM2] should be regarded as a
pioneering work).

A simpler example of a quantitative (yet quite abstract) consequence of
the above principle is the determination of the density function
mentioned in the principle: the latter is
in fact essentially determined.
If we are interested in stationary distributions
phenomena which are observable by measurements that take place in a
fixed time $\t$ we can just take averages over $A$ with respect to a
distribution with density over $A$ proportional to $\L_{\t'}^{-1}(x)$,
with $\t'=M'\t_0\gg\t$ (where the expansion rate is the jacobian
determinant of the transformation $S_{\t'}$ at $x$, \ie
$\L_{\t'}^{-1}(x)\=\prod_{-M'}^{M'} \L_{\t_0}^{-1}(S^j_{\t_0}x)$).  So
that two equal area elements of $A$ around $x$ and $y$ have a {\it relative
probability} of visit equal to $\L_{\t'}^{-1}(x)/\L^{-1}_{\t'}(y)$.

Of course $\t'$ cannot be taken too large: if $\t'$ is taken of the
order of the recurrence time the ratio becomes $1$. The natural upper
bound on $\t'$ has to be such that the cells in $U$ ending in the
considered area elements are still in a large number. This sets an upper
limit to the values of $\t$ for which the above remark applies.
\footnote{${}^{19}$}{\nota this means that the ratio between the linear
dimension of $U$ and the linear dimension of the cells has to be large
compared to the maximal linear expansion rate over the time $\t$, a
condition that can be expressed in terms of the largest Lyapunov
exponent.}

The example \equ(4.1) is very special.\footnote{${}^{20}$}{\nota
this is shown also by the fact that the operation $i$ mapping $x=(\V
p,\V q)$ to $ix=(-\V p,\V q)$ is such that $t\to ix(-t)$ is a solution
of the equation of motion if $t\to x(t)$ is such: a time reversal
symmetry. This has several implications, among which the properties that
both initial data $x$ and $ix$ evolve towards the same attractor $A$, in
the future, and to the attractor $iA$ in the past. In general $A$ and
$iA$ are different, except in the case $\V E=\V 0$ (because $A$ is the
full energy surface).}

It is however generalizable: many generalizations have already been
considered in the literature, [PH].  Still it should be stressed that
the models to which the above principle can be applied form a rather
small class of deterministic models.  It is not immediately clear how
it can be applied to stationary non equilibrium phenomena in which the
thermostat is realized in a different way, \eg by some stochastic
boundary conditions.  Nor it is obvious that the different thermostats
are physically equivalent.

In my opinion there is, also, some misunderstanding in the literature
about the fact that the set $A$ has zero measure (in the non equilibrium
cases this has been sometimes associated with the questions related to
irreversibility) and about the fact that $A$, regarded as a folded
surface on the equal energy manifold, has a fractal dimension (thereby
representing a "strange attractor"). Such facts may be quite misleading.
The above analysis shows that $A$ should be more conveniently regarded
as a smooth non fractal surface of dimension about $6N/2$: its fractal
dimension arises from the folding of $A$ on the surface of constant
energy (rising from $6N/2$ to about $6N$ if $\V E$ is small).

Furthermore in the assumption that the stochastic thermostats and the
gaussian thermostat (or other thermostats, [PH]) are equivalent one sees
clearly a problem related to attaching importance to the set $A$ as a
fractal with zero measure. In fact we expect that stochastic thermostats
lead to stationary distributions which have a density in phase space,
hence which cannot be concentrated on a set of $0$ measure.

The contradiction disappears if one thinks that, in a stationary state,
there may be several distributions which, in the limit as $N\to\io$,
become equivalent. A distribution concentrated on a set of zero measure
might well be equivalent to one distributed on the whole energy surface,
or on the whole phase space. A much simpler, but very familiar, example
of such a situation
is provided by the microcanonical distribution which is concentrated on
a set of zero measure, but it is equivalent (in the thermodynamic limit)
to the canonical distribution, which is concentrated on the whole phase
space.

Finally it should be clear that the problem of approach to stationarity
will show up exactly in the same terms as in the equilibrium cases. The
"ergodicity" assumptions above cannot in any way justify the use of the
distribution verifying the Ruelle principle: the time necessary for a
phase space point to visit the full set of cells building $A$ will be of
the order of magnitude of the recurrence time. And as in the equilibrium
cases we can expect that the rapidity of the approach to equilibrium is
rather due to the fact that we are interested only in very few
observables, and such observables have the same value in most of phase
space.

\*
I hope to have shown, or at least given arguments, that the point of
view, see for instance [Pl], whereby Boltzmann was a XIX century
physicist judged by his interpreters with XX century mathematical
standards is not exactly correct: today's way of thinking is not too
different from his and most problems the physicists had with his work
were problems with the understanding of his Physics and {\it not} of
his Mathematics, see also [L].  The misunderstandings about his ideas
are, in my opinion, largely due to the unwillingness of studying the
original publications and to the unfounded belief that they were
forwarded with fidelity by the reviewers that wrote about his
achievements.

\*
{\bf Acknowledgements:} I owe to my father Carlo essential help in the
explanation of the etymology of the word ergodic.  Part of the
interpretation of Ruelle's principle presented here was developed in
collaboration with E.  Cohen in a joint effort to understand more deeply
the results of the paper [ECM2]: while our analysis, which preceded this
paper, will be published elsewhere I wish to thank him for communicating
to me his enthousiasm on the subject while I was visiting Rockefeller
University, and for his thoughtful comments on this paper.  I am
indebted to J.  Lebowitz for his hospitality at Rutgers university and
for stimulating my interest on the gaussian thermostats.  To him I owe
also the redressement of several misconceptions and mathematical errors.

\pagina
\*
{\bf References.}
\*
\item{[B66]} Boltzmann, L.: {\it\"Uber die mechanische Bedeutung des
zweiten Haupsatzes der W\"armetheorie}, in
"Wissehschaftliche Abhandlungen", ed. F. Hasenh\"orl,
vol. I, p. 9--33, reprinted by Chelsea, New York).

\item{[B68]} Boltzmann, L.: {\it Studien \"uber das Gleichgewicht der
lebendigen Kraft zwischen bewegten materiellen Punkten}, in
"Wissehschaftliche Abhandlungen", ed. F. Hasenh\"orl,
vol. I, p.  49--96, reprinted by Chelsea, New York.

\item{[B72]} Boltzmann, L.: {\it Weitere Studien \"uber das
W\"armegleichgewicht unter Gasmolek\"ulen}, english translation in S.
Brush, {\it Kinetic theory}, {Vol. 2}, p. 88. Original in
"Wissehschaftliche Abhandlungen", ed. F. Hasenh\"orl,
vol. I, p. 316--402, reprinted by Chelsea, New York).

\item{[B77]} Boltzmann, L.: {\it \"Uber die Beziehung zwischen dem
zweiten Hauptsatze der mechanischen W\"armetheorie und der
Wahrscheinlichkeitsrechnung, respektive den S\"atzen \"uber das
W\"armegleichgewicht}, in "Wissenschaftliche Abhandlungen", vol. II,
p. 164--223, F. Hasen\"ohrl, Chelsea, New York, 1968 (reprint).

\item{[B84]} Boltzmann, L.: {\it \"Uber die eigenshaften monzyklischer
und anderer damit verwandter Systeme}, in "Wissenshafltliche
Abhandlungen", ed. F.P. Hasenh\"orl, vol. III,
Chelsea, New York, 1968, (reprint).

\item{[B96]} Boltzmann, L.: {\it Entgegnung auf die
w\"armetheoretischen Betrachtungen des Hrn.  E.  Zermelo}, english
translation in S.  Brush, "Kinetic Theory", {vol. 2}, 218--, Pergamon
Press.

\item{[B97]} Boltzmann, L.: {\it Zu Hrn.  Zermelo's Abhandlung "Ueber
die mechanische Erkl\"arung irreversibler Vorg\"ange}, english
translation in S.  Brush, "Kinetic Theory", {\bf 2}, 238.

\item{[B02]} Boltzmann, L.: {\it Lectures on gas theory}, english
edition annotated by S.  Brush, University of California Press,
Berkeley, 1964.

\item{[Ba]} Bach, A.: {\it Boltzmann's probability distribution of
1877}, Archive for the History of exact sciences, {\bf 41}, 1-40, 1990.

\item{[Br1]} Brush, S.: {\it The kind of motion we call heat}, North
Holland, 1976 (vol. II), 1986 (vol. I).

\item{[BSC]} Bunimovitch, L., Sinai, Y., Chernov, N: {\it Statistical
properties of two dimensional hyperbolic billiards}, Russian
Mathematical Surveys, {\bf 45}, n. 3, 105--152, 1990.

\item{[C]} Clausius, R.: {\it The nature of the motion which we call
heat}, in "Kinetic Theory, ed. S. Brush, p. 111---147.

\item{[CELS]} Chernov, K., Eyink, G., Lebowitz, J., Sinai, Y.:
{\it Steady state electric conductivity in the periodic Lorentz gas},
Communications in Mathematical Physics, {\bf 154}, 569--601, 1993.

\item{[D]} Dugas, R.: {\it La th\'eorie phisique au sens de Boltzmann},
Griffon, Neuch\^atel, 1959.

\item{[Dr]} Dressler, U.: {\it Symmetry property of the Lyapunov
exponents of a class of dissipative dynamical systems with viscous
damping}, Physical Review, {\bf 38A}, 2103--2109, 1988.

\item{[ECM1]} Evans, D.,Cohen, E., Morriss, G.: {\it Viscosity of a
simple fluid from its maximal Lyapunov exponents}, Physical Review,
{\bf 42A}, 5990--5997, 1990.

\item{[ECM2]} Evans, D.,Cohen, E., Morriss, G.: {\it Probability of second
law violations in shearing steady flows}, Physical Review
Letters, {\bf 71}, 2401--2404, 1993.

\item{[EE]} Ehrenfest, P., Ehrenfest, T.: {\it The conceptual
foundations of the statistical approach in Mechanics},
Dover, 1990, (reprint).

\item{[G]} Gibbs, J.: {\it Elementary principles in statistical
mechanics}, Ox Bow Press, 1981, (reprint).

\item{[G1]} Gallavotti, G.: {\it Aspetti della teoria ergodica
qualitativa e statistica del moto}, Quaderni dell' U.M.I., vol. 21,
ed. Pitagora, Bologna, 1982.

\item{[G2]} Gallavotti, G.: {\it L' hypoth\`ese ergodique et
Boltzmann}, in "Dictionnaire Phylosophique", Presses Universitaires de
France, p. 1081-- 1086, Paris, 1989.

\item{[G3]} Gallavotti, G.: {\it Meccanica Statistica}, entry for the
"Enciclopedia italiana delle scienze fisiche", preprint Roma, 1984.  In
print (scheduled publication, 1994).  The published version will also
include another entry, originally written to be a separate one, {\it
Equipartizione e critica della Meccanica Statistica Classica}, Roma,
preprint 1984.  See also the entry {\it Teoria Ergodica}, preprint
Roma, 1986, for the "Enciclopedia del Novecento", (in print? maybe).

\item{[G4]} Gallavotti, G.: {\it Insiemi statistici}, entry for the
"Enciclopedia italiana delle scienze fisiche", preprint Roma, 1984.
In print (scheduled publication, 1994).

\item{[He1]} Helmholtz, H.: {\it Principien der Statik monocyklischer
Systeme}, in "Wissenschaftliche Abhandlungen", vol.  III, p.  142--162
and p.  179-- 202, Leipzig, 1895.

\item{[He2]} Helmholtz, H.: {\it Studien zur Statik monocyklischer
Systeme}, in "Wissenschaftliche Abhandlungen", vol.  III, p.  163--172
and p.  173-- 178, Leipzig, 1895.

\item{[HHP]} Holian, B., Hoover, W., Posch. H.: {\it Resolution of
Loschmidts paradox: the origin of irreversible behaviour in reversible
atomistic dynamics}, Physical Review Letters, {\bf 59}, 10--13, 1987.

\item{[Ja]} Jacobs, K.: {\it Ergodic theory and combinatorics}, in
Proceedings of the conference on {\it Modern analysis ansd probability},
june 1982. Contemporary Mathematics, {\bf 26}, 171--187, 1984.

\item{[K]} Kuhn, T.: {\it Black body theory and the quantum
discontinuity.  1814--1912}, University of Chicago Press, 1987.

\item{[Kl1]} Klein, M.: {\it Maxwell and the beginning of the Quantum
Theory}, Archive for the history of exact
sciences, {\bf 1}, 459--479, 1962.

\item{[Kl2]} Klein, M.: {\it Mechanical explanations at the end of the
nineteenth century}, Centaurus, {\bf 17}, 58--82, 1972.

\item{[Kl3]} Klein, M.: {\it The development of Boltzmann statistical
ideas}, in "The Boltzmann equation", ed. E. Cohen, W. Thirring, Acta
Physica Austriaca, suppl. X, Wien, p. 53--106.

\item{[L]} Lebowitz, J.: {\it Boltzmann's entropy and time's arrow},
Physics Today, Sept 1993, p. 32--38.

\item{[LPR]} Livi, R., Politi, A., Ruffo, S.: {\it Distribution of
characteristic exponents in the thermodynamic limit}, Journal of
Physics, {\bf 19A}, 2033--2040, 1986.

\item{[M]} Maxwell, J.: {\it On Boltzmann's theorem on the average
distribution of energy in a system of material points}, in "The
scientific papers of J,C, Maxwell", ed. W. Niven, Cambridge University
Press, 1890, vol. II, p. 713--741.

\item{[Ma]} Mathieu, M.: {\it On the origin of the notion 'Ergodic
Theory'}, Expositiones Mathematicae, {\bf 6}, 373--377, 1988. {\it See
footnote ${}^9$ above}.

\item{[Pl]} Plato, J.: {\it Boltzmann's ergodic hypothesis},
Archive for the History of exact sciences, {\bf 44}, 71-89, 1992.

\item{[H]} Posch, H., Hoover, W.: {\it Non equilibrium molecular
dynamics of a classical fluid}, in "Molecular Liquids: new perspectives
in Physics and chemistry", ed. J. Teixeira-Dias, Kluwer Academic
Publishers, p. 527--547, 1992.

\item{[R]} Ruelle, D.: {\it Measures describing a turbulent flow},
Annals of the New York Academy of Sciences, {\bf 357}, 1--9, 1980.
See also Eckmann, J., Ruelle, D.: {\it Ergodic theory of strange
attractors}, Reviews of Modern Physics, {\bf 57}, 617--656, 1985; and
Ruelle, D.: {\it Ergodic theory of differentiable dynamical systems},
Publications Math\'emathiques de l' IHES, {\bf 50}, 275--306, 1980.

\item{[S]} Sinai, Y.: {\it Dynamical systems with elastic reflections.
Ergodic properties of dispersing billards},
Russian Mathematical Surveys, {\bf 25}, 137--189, 1970.

\item{[Sc]} Schwartz, J.: {\it The Pernicious Influence of Mathematics on
Science}, in "Discrete thoughts: essays in Mathematics, Science,
and Phylosophy", M. Kac, G. Rota, and J. Schwartz, eds., Birkhauser, Boston,
1986, p. 19--25.

\item{[SEM]} Sarman, S., Evans, D., Morriss, G.: {\it Conjugate pairing
rule and thermal transport coefficients}, Physical Review, {\bf 45A},
2233--2242, 1992.

\ciao